# Probabilistic modeling reveals coordinated social interaction states and their multisensory bases

**Authors:** Sarah Josephine Stednitz[1]*, Andrew Lesak[2]*, Adeline L Fecker[2], Peregrine Painter[2], Phil Washbourne[2]+, Luca Mazzucato[2]+, Ethan K Scott[13]+

*Joint first authors    +Joint last authors

**Affiliations:**
1.) Department of Anatomy & Physiology, University of Melbourne, Parkville, VIC, Australia
2.) Institute of Neuroscience, University of Oregon, Eugene, OR, USA
3.) Queensland Brain Institute, University of Queensland, St Lucia, QLD, Australia

## HIGHLIGHTS
- Zebrafish exhibit distinct correlated interaction states with unique timescales.
- Delayed interactions are visual while synchronization requires mechanosensation.
- A new class of hidden Markov model segments social interactions into discrete states.
- States alternate within a session, revealing real-time dynamics of social behavior.

## ABSTRACT

Social behavior across animal species ranges from simple pairwise interactions to thousands of individuals coordinating goal-directed movements. Regardless of the scale, these interactions are governed by the interplay between multimodal sensory information and the internal state of each animal. Here, we investigate how animals use multiple sensory modalities to guide social behavior in the highly social zebrafish (*Danio rerio*) and uncover the complex features of pairwise interactions early in development. To identify distinct behaviors and understand how they vary over time, we developed a new hidden Markov model with constrained linear-model emissions to automatically classify states of coordinated interaction, using the movements of one animal to predict those of another. We discovered that social behaviors alternate between two interaction states within a single experimental session, distinguished by unique movements and timescales. Long-range interactions, akin to shoaling, rely on vision, while mechanosensation underlies rapid synchronized movements and parallel swimming, precursors of schooling. Altogether, we observe spontaneous interactions in pairs of fish, develop novel hidden Markov modeling to reveal two fundamental interaction modes, and identify the sensory systems involved in each. Our modeling approach to pairwise social interactions has broad applicability to a wide variety of naturalistic behaviors and species and solves the challenge of detecting transient couplings between quasi-periodic time series.

## INTRODUCTION
Animals must actively perceive conspecific cues to generate context-appropriate behavior. Selecting the optimal response to a social cue enhances survival and reproductive fitness, whether by avoiding predation, increasing access to food, or courting prospective mates. Such interactions can consist of coordinated actions between individuals, and often involve multiple sensory systems. It is critical to attend to the correct sensory modality for a given context - for example, it may be advantageous to watch a rival's movement during physical conflict, while disregarding distracting information such as their scent. In other cases, multiple sensory modalities form a holistic stimulus, as in a conversation where the voice of a social partner and their facial expressions are both essential for the correct interpretation. The relevant sensory stimuli differ across vertebrate species depending on the ethological and environmental context. Teleost fish experience a considerably different sensory environment than many terrestrial animals - they navigate freely in three dimensions; experience changes in water flow and pressure, and encounter complex chemosensory signals from prey, predators, and conspecifics. Despite these differences, they fundamentally process the same types of sensory information as mammals and other animals when navigating their social environment: sight, sound, touch, and smell. Do these sensory modalities operate independently to evoke behavior, or do they coalesce into a combined perception of a stimulus?

We leverage the teleost zebrafish (*Danio rerio*) to interrogate how multimodal sensory information underlies coordinated social behaviors. Zebrafish are a highly social species in the wild, aggregating in groups of up to 300 individuals, and occupy diverse habitats across southeast Asia (Suriyampola et al., 2016, Engeszer et al., 2007). Larval zebrafish are precociously independent and attracted to conspecifics early in development, typically by 14-21 days post-fertilization (dpf). Rudimentary interactions such as preferential turns develop first, and later increase in complexity to form the full repertoire of adult behaviors (Dreosti et al 2015, Harpaz et al 2021, Hinz & de Polavieja 2017, Stednitz & Washbourne 2020). This comparatively rapid developmental progression is particularly interesting in the context of early circuit formation. As a vertebrate, much of the genome is conserved with humans, and despite superficial differences in neuroanatomy, many underlying processes and structures are conserved across species. Indeed, some of the anatomical regions implicated in the brain's social decision-making network are found in both mammals and teleosts, to the extent that molecular and functional identities are maintained (Mueller & Wulliman 2009, O'Connell & Hofmann, 2012) and neurons in these areas modulate their activity in social contexts (Kappel et al 2022, Pinho et al 2023, Teles et al 2015). This shared circuitry represents a unique opportunity to probe the mechanisms underlying social development across animal species. A functional circuit approach, however, depends on the reliable classification of naturalistic interactions and the identification of sensory cues that optimally evoke these behaviors.

Although early-stage larvae do not aggregate, fish reared in isolation are hypersensitive and startle more often in social contexts, suggesting that mechanosensory cues produced by other larvae are necessary for normal sensory processing and social development (Groneberg et al., 2020). Socially isolated larvae also exhibit both rapid neuromodulatory changes in oxytocin signaling and the production of vertebrate-specific neuropeptides such as *pth2* (Wee et al, 2022; Anneser et al., 2020). Interestingly, these effects are modulated by different sensory modalities, with olfaction inducing oxytocin signaling and mechanosensation underlying *pth2* regulation. Indeed, virtual assays that employ purely visual stimuli are effective at eliciting some aspects of social interaction, but do not evoke the full suite of behaviors or brain activity as compared to the rich multisensory stimulus generated by a real conspecific (Larsch & Baier 2018, Kappel et al 2022). Despite this, purely visual features of conspecifics like body form and color patterns drive aggregation, and biological motion is also sufficient to induce attraction in both juvenile and adult fish (Larsch & Baier 2018, Nunes AR 2020, Polverino et al 2012, Rosenthal & Ryan 2005, Saverino & Gerlai 2008).

Teleost mechanosensation is achieved in part via the lateral line, a series of sensory hair cells along the body and head of the animal that are deflected by small changes in water flow or pressure. This dynamic sensory input provides important social cues, allowing fish to coordinate movements with animals that are outside the field of view, or avoid threats or obstacles detected by others (Mekdara et al 2018, Partridge BL & Pitcher TJ 1980). Flow sensation facilitates alignment and synchronized tail beats in schooling fish, and promotes short-range attraction even in the absence of visual cues (Lombana & Porfiri 2022, Tidswell et al 2024). Given that zebrafish are highly dependent on both vision and mechanosensation to navigate their natural environment, a combination of sensory modalities could be important for optimal social behavior, and there is an intriguing potential to investigate whether multisensory integration is required to generate higher-order behaviors.

The classification of social behavior presents several challenges, even when considering simplified pairwise interactions. First, the behavior of one animal serves as a stimulus to the other, resulting in positive feedback that requires the analysis of both animals to fully describe the sensory input and behavioral output (Katz et al 2011, Stowers et al 2017; Stednitz et al 2018). These actions, feedback, and responses occur on relatively brief time scales, but interactions also vary over time such that animals engage and disengage across seconds, minutes, hours, and days, governed by both the environment and the internal states of individuals (Calhoun et al, 2019, Laan et al 2018, Sridhar VH 2023, Wang et al 2024, Xue et al 2023). Phase transitions between different modes of collective behavior are dependent on the actions of individuals, each experiencing their own unique sensory information (Guttal et al 2012, Poel et al 2022). This continuous state-dependent variability renders it particularly challenging to measure interactions using summary statistics over a given recording period. Therefore, a rich description of social behavior depends on capturing both the coordination of multiple animals and the system's long-term dynamics. Juvenile zebrafish are particularly well suited to this type of time series analysis, as they perform discrete movement bouts that are easily captured and segmented, rather than the continuous swimming exhibited by adults. Consequently, both the stimulus (the social partner) and the response (the behavior of the focal fish) can be readily measured and their relationships described. The behavior of both animals is important to consider, given that reciprocal coupling of swim bouts between fish is necessary to describe the emergent properties of movement patterns between pairs (Amichay et al 2024). However, there remains the possibility that different timescales of interaction occur naturally, vary over time, and may depend on unique sensory modalities.

Hidden Markov models (HMMs) are an attractive choice for the probabilistic segmentation of behavioral data over time because they can agnostically reveal features in continuous time-series data corresponding to distinct behavioral motifs (Calhoun 2019, Costacurta et al 2022, Johnson et al 2019, Markowitz et al 2018, Parker et al 2022, Mazzucato 2022). However, their usefulness depends on the careful selection of training data that encapsulates key features intended for identification, such as pose tracking data or keypoints (Weinreb et al 2024). Further, changes to the environment or context may result in slight variations on the behavioral states expressed in a given pair, and fitting an HMM with a large number of parameters for many pairs on an individual basis can be computationally demanding and vulnerable to overfitting and noise (Lee et al 2023). A specific feature of social interaction is the inherent periodicity of single fish movement bouts, which poses an unresolved challenge to modeling efforts aimed at uncovering transient epochs where quasi-periodic signals are coupled. In particular, state space models relying on the joint probabilities of the two time series are sensitive to the presence of spurious or "nonsense" correlations in this regime (Harris 2021, Yule 1926).

To overcome these challenges and better understand the role of multiple sensory signals in the early ontogeny of social behavior, we studied the pairwise interactions of zebrafish in early development within a naturalistic context. We then developed a constrained linear-model hidden Markov model (cLM-HMM) to classify behavioral states by using the movement of one animal to predict that of its partner. The introduction of constraints on model parameters allowed a robust and efficient identification of three interaction states: two social states characterized by fast synchronization or delayed interactions, and a non-interacting state. We then removed input from the visual and/or mechanosensory systems to uncover specific relationships between modalities and interaction modes, revealing that in-phase synchronized and out-of-phase delayed interaction states are driven by mechanosensory and visual cues, respectively, thus providing a framework for understanding the sensorimotor landscape underlying social interactions. Importantly, our model demonstrates a general framework in which unique interaction states can be identified for systems that exhibit switching correlated structure between multiple variables.

## RESULTS
### SPATIAL PREFERENCE & COORDINATION

We first recorded size-matched sibling pairs of larval zebrafish (at 11-13 dpf) in a 50mm circular arena to determine the earliest detectable onset of social interaction in our assay (Fig 1A). We measured the body angle, distance from the partner, and the estimated field of view for each fish (Fig 1B-C), and found that juvenile zebrafish begin to exhibit early forms of shoaling by maintaining close distances to conspecifics at 12-13 dpf (Fig 1E-F), when larvae are 6.7 +/- 1.4 mm in length (Fig S1A), corresponding to the early flexion stage (Parichy et al, 2009). We found that pairs favor smaller median distances and that this preference increases with age, with significantly closer-than-random proximity by 12 dpf (ANOVA main effect for age, p =

0.036, Fig 1E). We created surrogate datasets with pseudosessions obtained by randomly swapping one fish in each pair with a fish from another (Fig 1D), confirming that our observed spatial relationships are not spurious, as their median distances are smaller than those in our shuffled data (p = 0.013, Fig 1E). This comparison with surrogate shuffled datasets will be used throughout to control for spurious correlations that may arise due to spatial preferences or single fish movement statistics (Harris 2021, Yule 1926).

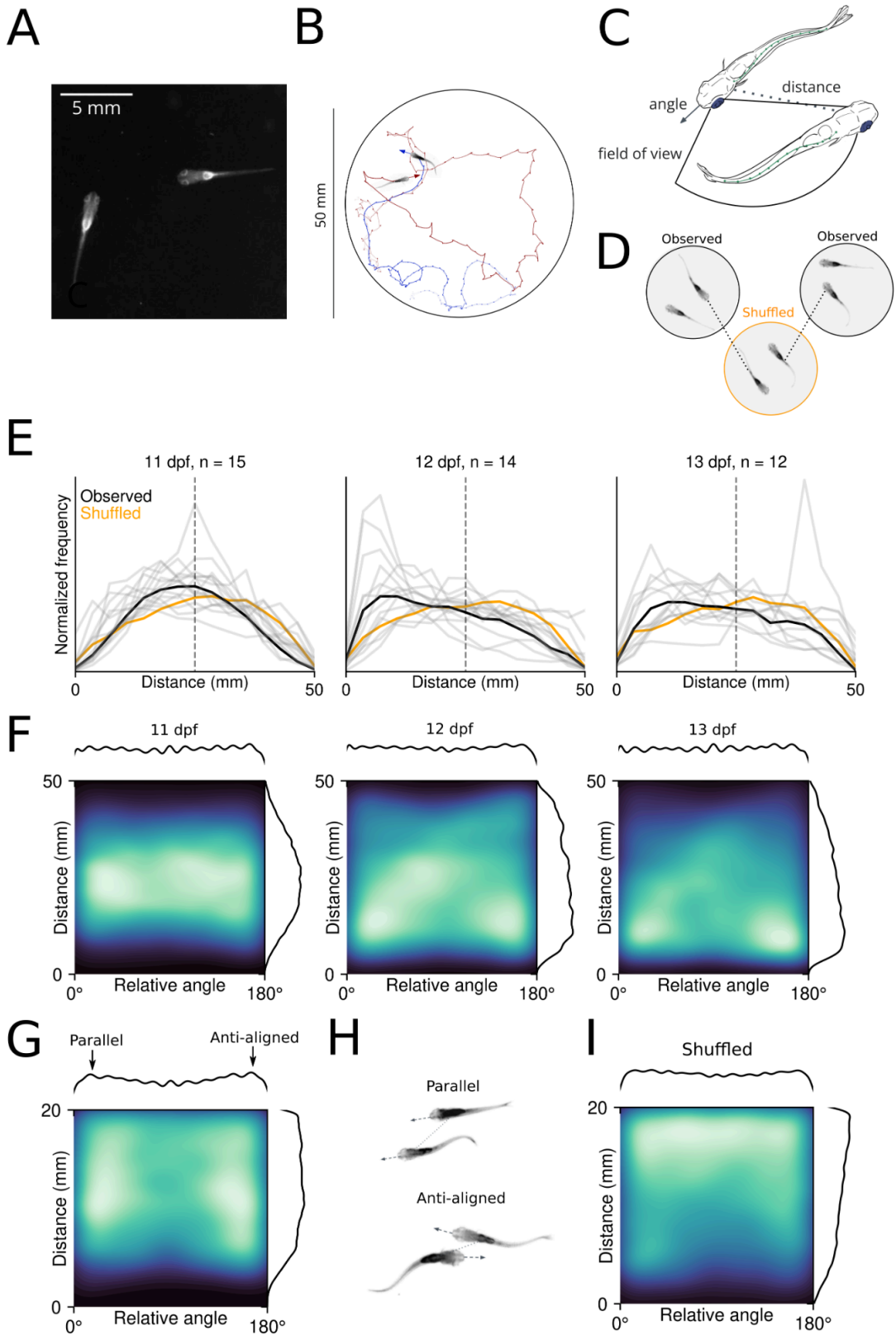

**Figure 1. A.)** Example image of a pair of juvenile zebrafish as recorded in our assay. **B.)** Example trajectories and headings of two fish in a 50 mm circular arena. **C.)** Behavioral variables measured following tracking: the heading angle of each fish, the angular position of the other fish relative to the target fish's heading, and the distance between the two fish. **D.)** Schematic of shuffling operation, where real data are swapped between experiments to control for spurious relationships. **E.)** Histograms of preferred distances across development each pair (gray lines), with the mean indicated in black. **F.)** 2D histograms of distance and relative angle across development, showing preferred orientations at close distances. **G.)** 2D histograms for 12-13 dpf at <20 mm distance, arrows on marginal plot above indicate parallel and anti-aligned relative headings. **H.)** Schematic of parallel and anti-aligned configurations indicated by high density regions in Figure G. **I.)** 2D histograms of shuffled data at <20 mm distance.

Juvenile zebrafish adopt specific orientations relative to conspecifics, and are generally aligned (parallel) or anti-aligned (facing opposite directions) to their partner (Fig 1F). Animals are most frequently found at these orientations when they are in close proximity (Fig 1G). We hypothesize that these orientations reflect distinct behaviors: schooling-like polarized swimming (with fish parallel) and visually-mediated shoaling and orienting (anti-aligned) (Fig 1H). In contrast, such structure is absent from shuffled data, and real pairs are statistically more likely to adopt these positions than shuffled pairs (Fig 1I, Fig S1B). Importantly, we observe the same configurations in adult animals recorded in similar conditions, suggesting that these preferences develop early in juvenile stages and are maintained throughout the life of the animal (Fig S1C).

Teleosts maintain visual contact with conspecifics during schooling, and the number of fish in the visual field guides both simple orienting and complex dominance-mediated interactions (Davidson et al 2021, Harpaz et al 2021, Rodriguez-Santiago et al, 2020). The geometry of the eye position constrains the field of view to the sides of the animal, with the exception of during prey capture events where the eyes are temporarily converged to provide depth information (Bianco et al, 2011). We calculated the position of the partner fish's body relative to the head of the focal fish, assuming that 10 degrees immediately in front of and behind the fish are blind spots (Fig 2A, Pita et al 2015). The partner fish is less likely to be located in these blind spots as compared to shuffled data (p = 0.005), consistent with the idea that zebrafish maintain configurations where partners are visible (Fig 2B).

Social behavior is increasingly coordinated over development (Dreosti et al 2015, Stednitz & Washbourne 2020), making it likely that fish would coordinate their movements with one another. Movement bouts are characterized by bursts of high-velocity swim events (Fig 2C), and bouts are inherently periodic in isolated fish, as revealed by the autocorrelograms of the single-fish speeds (Fig 2D). The fish in our experiment executed a swim bout on average every 0.6 seconds, with a linear decrease in the interbout interval over development ($R^2$ = .137, p < 0.001, Fig S1D). The cross-correlograms between the speeds of paired fish revealed that the movements of one animal depend on its partner across time and are correlated in a periodic structure (Fig 2E). To exclude the possibility that the cross-correlation feature between the animal speeds were solely determined by the inherent periodicity of individual fishs' movements rather than social interaction, we performed a control analysis where we estimated cross-correlograms in the surrogate dataset with shuffled fish pairs (see Fig 1D). The periodicity in the cross-correlogram is only present in the real data and abolished in the surrogate datasets (Fig 2F), and interbout intervals are significantly correlated within real data but not shuffled pairs ($R^2$ = .321, p < 0.001 and $R^2$ = .002, p = 0.739 respectively, Fig S1E). Instantaneous correlations between movements are more tightly coupled when animals are physically close, while delayed components are evident at greater distances (Fig 2G-H). We thus concluded that the relationship between interbout intervals of the two fish is not due to the inherent single-fish periodicity in movements, but rather reflects coordination between individuals.

Overall, our observations and analyses of pairs of juveniles show that, starting at roughly 12 dpf, fish begin to swim in close proximity to one another, coordinate their movements, and align themselves in parallel or antiparallel orientations. The observation that movements are more tightly coordinated when fish are closer to one another may simply be a function of better sensory information at shorter distances, or may represent distinct modes of short- or long-range interactions.

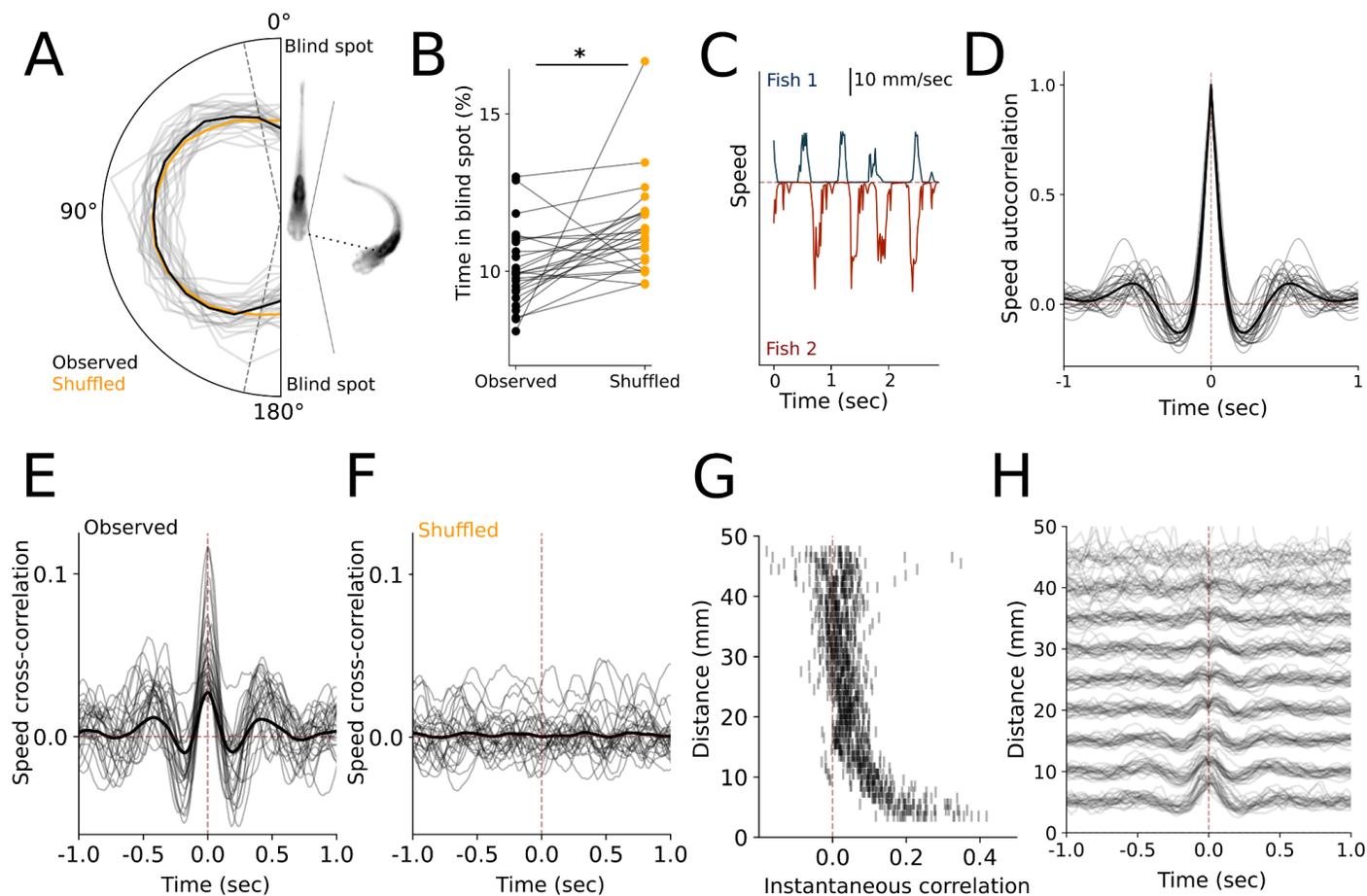

**Figure 2. A.)** Histogram of the partner position relative to the heading of the target fish, with the average indicated in black. **B.)** The percentage of time spent in the estimated blind spots in real vs shuffled pairs. **C.)** Example speeds of two interacting fish. **D.)** Speed autocorrelogram for all individual fish, showing periodic movements. **E.)** Cross-correlogram of fish speed between members of observed pairs. **F.)** Cross-correlogram of fish speed between members of shuffled pairs. **G.)** Instantaneous cross-correlation of partner fish speeds as a function of distance. **H.)** Relative comparison of cross-correlogram amplitudes as a function of distance.

## MULTIPLE SOCIAL STATES

Based on differences in distance and timing, we hypothesized that pairs of animals transition between different modes of interaction over time, and the temporal profile of these epochs may differ depending on the dominant sensory modality employed. While the average cross-correlogram measure presented above shows that fish interact by coordinating their swim bouts, this observation across the entire experiment only summarizes global activity and fails to capture the fine-grained temporal dynamics of their interactions. Studying the dynamics of such a system requires the classification of distinct behavioral motifs from one moment to the next. To achieve this, we employed a dynamical modeling approach to identify epochs when pairs are likely engaged in interactions, thereby uncovering the temporal features of their behavior.

We found the strongest indicator of social interaction to be the structure of coordinated bout sequences represented by bursts of high-velocity swim events (Fig 3A). Indeed, the latencies between either animal's next bout (referred to as "relative interbout intervals") are highly correlated, as captured by the periodic structure in the cross-correlogram, suggesting that coordinated movement between the fish persists for at least a few seconds and exhibits both a synchronized and a delayed component (Fig 3B). The distribution of relative interbout intervals is bimodal, unlike the unimodal latencies to the animal's own next bout (Fig 3C). This finding led us to hypothesize that the fast and slow modes of this distribution represent two distinct states of interaction. To more closely investigate the interdependency of each fish's movements, we examined the phase response curve (PRC) capturing the relationship between an individual's interbout interval duration and its latency from the partner fish's last bout (Amichay et al, 2024). The PRC exhibited a multi-modal distribution, suggesting the presence of two distinct interaction states: a synchronized, in-phase response state and a delayed out-of-phase response state (Fig 3D). Consistent with our prediction that these correlations represent social interactions, the multi-modal structure was absent in the shuffled dataset, which instead reflects the refractory period of a fish's own movements (Fig 3D).

We further reasoned that fish pairs may deploy these distinct interaction states at different times within each session. To uncover the dynamics of these distinct interaction states and their specific temporal features, we introduced a model based on two simple ingredients. The first ingredient is based on the observation that the cross-correlogram of the pair velocities, encoding social interactions, is mathematically equivalent to a linear model. In other words, we predict the velocity of one fish $v_2(t)$ given the velocity of the other fish $v_1(t)$ as:

$$v_2(t) = [W \star v_1](t) + b \tag{1}$$

where the bias $b$ is a constant offset and $\star$ represents the linear operation of temporal convolution between the input $v_1$ and a set of weights $W$ (see Methods). The cross-correlation $R_{12}(\tau)$ between $v_1(t)$ and $v_2(t)$ can be readily expressed as $R_{12}(\tau) = \frac{\sigma_1}{\sigma_2}[W \star R_1](\tau)$, where $R_1(\tau)$ is the autocorrelation of $v_1(t)$ and $\frac{\sigma_1}{\sigma_2}$ denotes the ratio of standard deviations between the respective fish speeds (Fig 2E, see supplementary for details). The second ingredient is the introduction of multiple such linear models each representing a different interaction state with a dynamical rule to determine how the pair switches between states on a moment-to-moment basis. This is achieved by modeling the temporal evolution of the interaction states as a Markov chain with transition matrix $A_{kl} = p(z_{t+1} = k | z_t = l)$ for states $k = 1,...,K$, encoding the probability of switching from state $l$ at time $t$ to state $k$ at time $t + 1$.

These two ingredients define a hidden Markov model with linear model emissions (Fig 3E), where each hidden state is associated with a unique linear model containing weights $W^{(k)}$ and biases $b^{(k)}$ (Fig 3E-F, Cazettes et al 2023). To construct this model, we let $v_1(n)$ and $v_2(n)$ denote the speeds of the input and focal fish, respectively, for $n = 1,...,T + 2L$ and structure the $D = 2L + 1$ dimensional input $\vec{v}_1(t) = [v_1(t), ..., v_1(t + 2L)]^T$ and output $\vec{v}_2(t) = [v_2(t), ..., v_2(t + 2L)]^T$ vectors to be the symmetrically time-lagged speeds of the input and focal fish, respectively, about time point $t + L$ for $t = 1,...,T$. In order to reliably detect sustained interaction states instead of transient correlations that emerge by chance in strongly periodic signals, the lag $L$ should be chosen to be longer than the autocorrelation time of each fish but shorter than the duration of social interactions as determined by the correlograms (Fig 3B). For our system, a lag of $L = 2$ seconds is sufficient. We express the linear model as:

$$\vec{v}_2(t) = W^{(k)} \vec{v}_1(t) + b^{(k)} \vec{1} + \epsilon(t) \tag{2}$$

where $\epsilon(t)$ is i.i.d Gaussian noise with zero mean and spherical covariance $\Sigma = \sigma^2 I$ tied across all states. The weights matrix $W^{(k)}$ in the hidden state $k$ captures a specific mode of social interaction, leading to $K$ state-specific cross-correlograms

$$R_{21}^{(k)}(\tau) = \frac{\sigma_1^{(k)}}{\sigma_2^{(k)}}[\vec{w}_L^{(k)} \star R_1^{(k)}](\tau) \tag{3}$$

where $\vec{w}_L^{(k)}$ is the center row of the weights matrix, $R_1^{(k)}(\tau)$ and $R_{21}^{(k)}(\tau)$ denote the state-conditioned autocorrelation and cross-correlation, respectively, and $\frac{\sigma_1^{(k)}}{\sigma_2^{(k)}}$ is the ratio of the state-conditioned standard deviations of the respective fish speeds (Fig 3E-G, see Methods & Supplementary for full details). The last technical ingredient in the model is that we constrain the weights matrix $W^{(k)}$ to have a symmetric Toeplitz structure, whereby each descending diagonal from left to right is constant. The Toeplitz structure naturally arises from the construction of the symmetrically time-lagged input and output vectors. The symmetric structure results from simultaneously fitting our model to all fish pairs while considering both permutations of the input and focal fish. The benefit of introducing a constrained weight matrix is to drastically reduce the number of free parameters from $D^2$ to $D$, enabling fast and efficient inference. We refer to the resulting model as a constrained linear-model HMM (cLM-HMM). Using these weights, the state-specific cross-correlations of pairs can be predicted by convolving the autocorrelation of the single-fish speed with the center row of the Toeplitz weight in each state (Fig 3G).

Consistent with the distribution of relative interbout intervals (Fig 3C-D), we identified two interaction states corresponding to fast- and slow-responding interactions, representing an in-phase synchronized and an out-of-phase delayed response, respectively. We further identified a null state where fish do not interact, yielding a total of three hidden states. We thus fit a three-state cLM-HMM globally to all pairs and obtained the weight matrices $W^{(k)}$ and biases $b^{(k)}$ for each interaction state (Fig. 3F). The HMM fits provide the posterior probability of each state being expressed for each time point, from which we assign the most likely state to each frame that exceeds the probability threshold of 0.8. This procedure allows us to revisit the dependence of relative angles (Fig. 1H) and cross-correlograms (Fig. 2GH) on distance by segmenting the time course of every session into discrete epochs, each one expressing a single interaction state, to reveal how pairs switch among the three states within a single session (Fig 4A-B).

State-conditioned distance and relative angle measurements reveal that synchronized interactions occur at short range in parallel/anti-aligned configurations, and delayed interactions occur at medium range while parallel (Fig 4C). Importantly, no significant states were identified when the cLM-HMM was fit to the shuffled data, as evidenced by the small values and lack of structure in the resultant weight matrices. This further confirms that the weights derived from the empirical data are not a result of spurious correlations (Fig S2A).

The cLM-HMM results shed light on the multi-modal nature of the phase-response curve in (Fig. 3D), revealing a state-specific relationship between relative interbout intervals, indicated by low latency in-phase interactions in the synchronized state and linearly correlated latencies in both the synchronized and delayed states (Fig 4D). To understand whether this linear relationship occurs because individual pairs have preferred interbout intervals, or whether the bout timings vary over the course of the experiment, we calculated the average interbout interval for a representative pair of fish across the three states. We found that interbout intervals are not stationary throughout the duration of a recording, and are therefore not an inherent property of individual animals but rather adjusts over time (Fig 4E). The null state, conversely, does not exhibit these linear relationships between interbout intervals for a given pair. While all three interaction states are long-lived (median durations of 9.7, 9.1, and 6.3 seconds for synchronized, delayed, and null state, respectively), synchronized interactions are sustained over a slightly longer duration than delayed interactions (Fig 4F). The switching dynamics further exhibited state-specific features where the two interacting states alternate with roughly equal probability, while the null state is most likely to transition to the delayed state (Fig 4G). This suggests that under normal conditions the null state occurs only briefly before animals resume coordinated swimming.

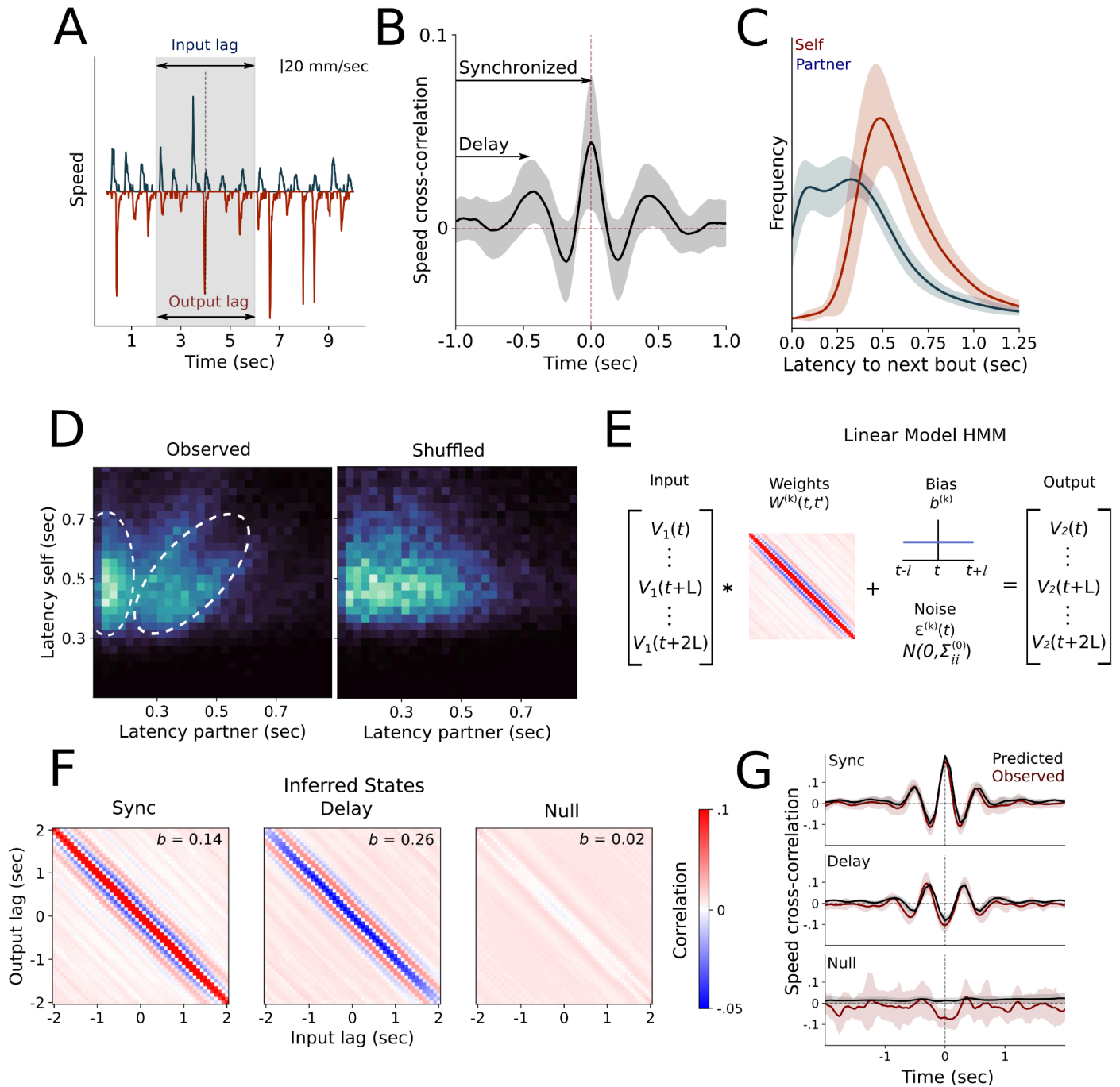

**Figure 3. A.)** Example velocities of an interacting pair, with the +/- 2 second input and output lags highlighted in the shaded area. **B.)** Time-lag velocity cross-correlation, with synchronized and delay elements indicated. Shaded area represents SEM. **C.)** Histograms of latency to both the target fish's own next bout, and the partner's next bout. **D.)** Phase response curves represented as 2-dimensional heatmap of latency to target fish's own next bout, and the partner's next bout. White dotted lines indicate both in-phase and out-of-phase components of interbout interval coordination in the observed data. **E.)** Schematic of the cLM-HMM. **F.)** Inferred states after fitting our constrained cLM-HMM to the full behavioral dataset. **G.)** Observed and predicted cross-correlograms for a representative pair, obtained by convolving the central row of the weights matrix in panel F with the state-weighed autocorrelation of one fish, scaled by the ratio of state-weighed standard deviations of each fish's speed (see Supplementary Material for details).

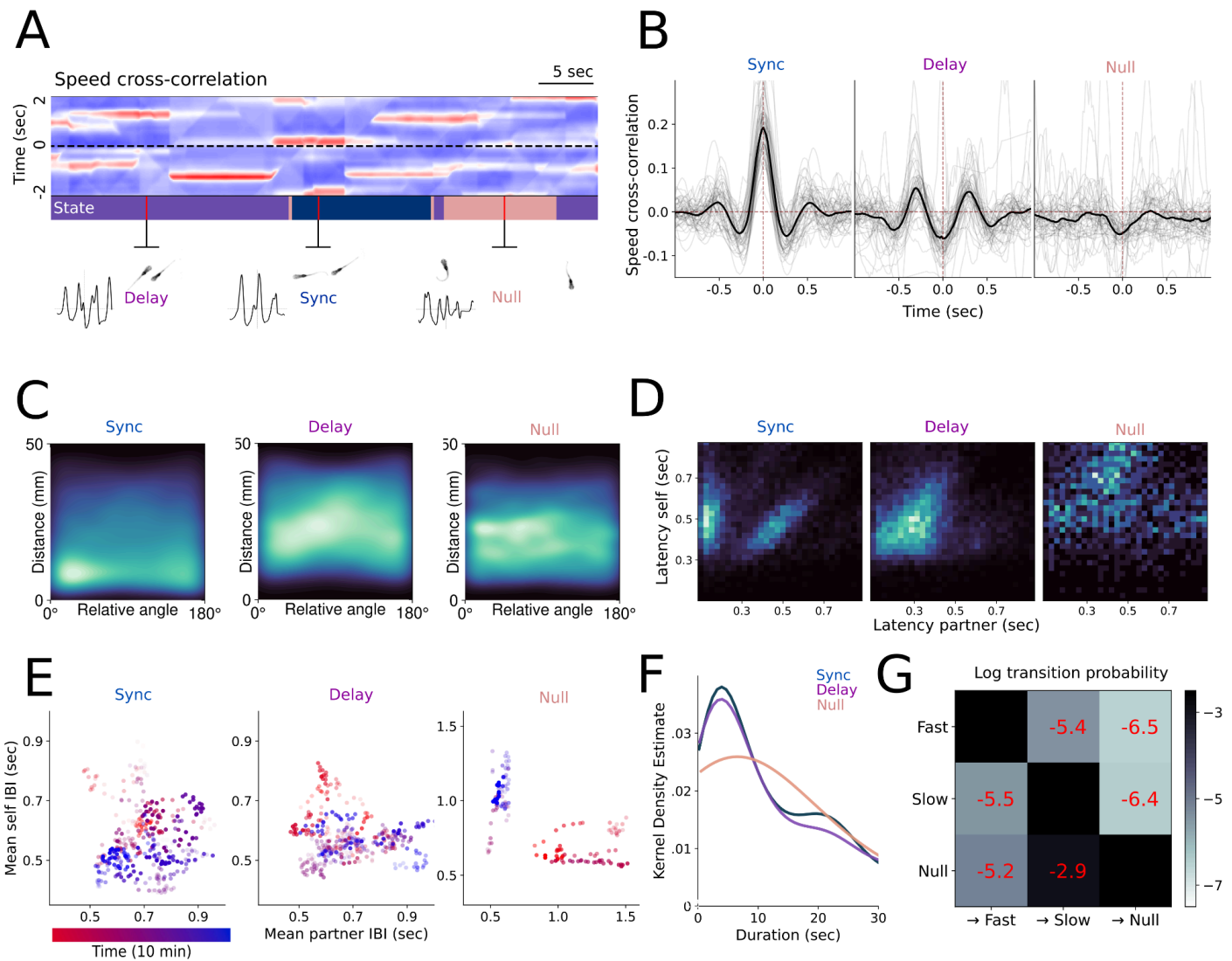

**Figure 4.**
**A.)** Example cross-correlations over a 2s sliding window, with the states inferred from the cross-correlations above annotated across time. **B.)** Cross-correlations derived from behavioral data after segmentation into discrete states. Grey lines are individual pairs, while the black line represents the mean. **C.)** 2D histograms of distance and relative angle following state segmentation, showing distance and alignment preferences for each state. **D.)** Relative interbout interval coordination across all states represented by phase response curves, indicating in-phase and out-of-phase components in the two interaction states. **E.)** Average interbout intervals for each state for a representative interacting pair over a sliding window of 1 minute, with time indicated by color, showing that preferred interbout intervals change over the course of the experiment. Opacity of data points corresponds to the correlation between all interbout intervals over the sliding window. **F.)** Kernel density estimate of state duration distribution. **G.)** Lexical transition matrix between states represented in log scale.

## SENSORY CUES

To investigate how ethologically relevant sensory modalities uniquely contribute to these specific states of social interaction, we ran an additional set of experiments to manipulate visual and mechanosensory cues. We reasoned that these sensory modalities operate on distinct time scales, and could therefore subserve different modes of interaction. We recorded interactions under both white light and dark conditions to understand the contribution of visual cues, and similarly disrupted the dominant form of flow sensation in zebrafish (Vanwalleghem et al, 2020) by ablating neuromasts of the lateral line system via neomycin exposure prior to the experiment (Fig 5A). We tested four possible combinations: light, dark, neomycin-treated in the light (neo-light), and neomycin-treated in the dark (neo-dark) (Fig 5B). We found that the proximity-driven orienting behaviors are altered under these sensory conditions (Fig 5C) and note similar patterns in adult pairs of zebrafish (Fig S2B). In particular, animals in the dark still maintain close, parallel swimming, with a reduced anti-aligned component. In contrast, neomycin-treated animals are found at greater distances and more likely to be anti- or orthogonally aligned to social partners (Fig 5C, S2B), and the median distance of neo-light pairs is statistically indistinguishable from our shuffled dataset (p = 0.834, Fig 5C, S2C). Interestingly, neo-dark pairs maintained closer-than-chance distances (p = 0.028), suggesting that our sensory blockade was incomplete or that another sensory modality was filling in for lost vision or water flow perception.

    The overall coordination between swim bouts was largely unaffected in animals interacting in the dark, while coordination is suppressed in lateral line ablated animals (Fig 5D). This observation, coupled with the presence of spatial preference even in the dark, indicates that non-visual cues can drive affiliative behavior (Fig 4C, S2C). The reduced coordination in lateral line ablated animals is not entirely accounted for by the increased distance between animals, as indicated by a suppressed instantaneous correlation relative to intact pairs even when the animals are in close proximity (Fig S2E). Despite the diminished correlation, we find that lateral line ablated pairs still tend to travel similar distances as their partner fish (Fig S2F), and that movements are broadly suppressed in the dark. We observe low latency bout coordination in all conditions with the exception of neo-dark, and a suppression of the linear relationship between delayed interbout intervals in ablated pairs (Fig 5E).

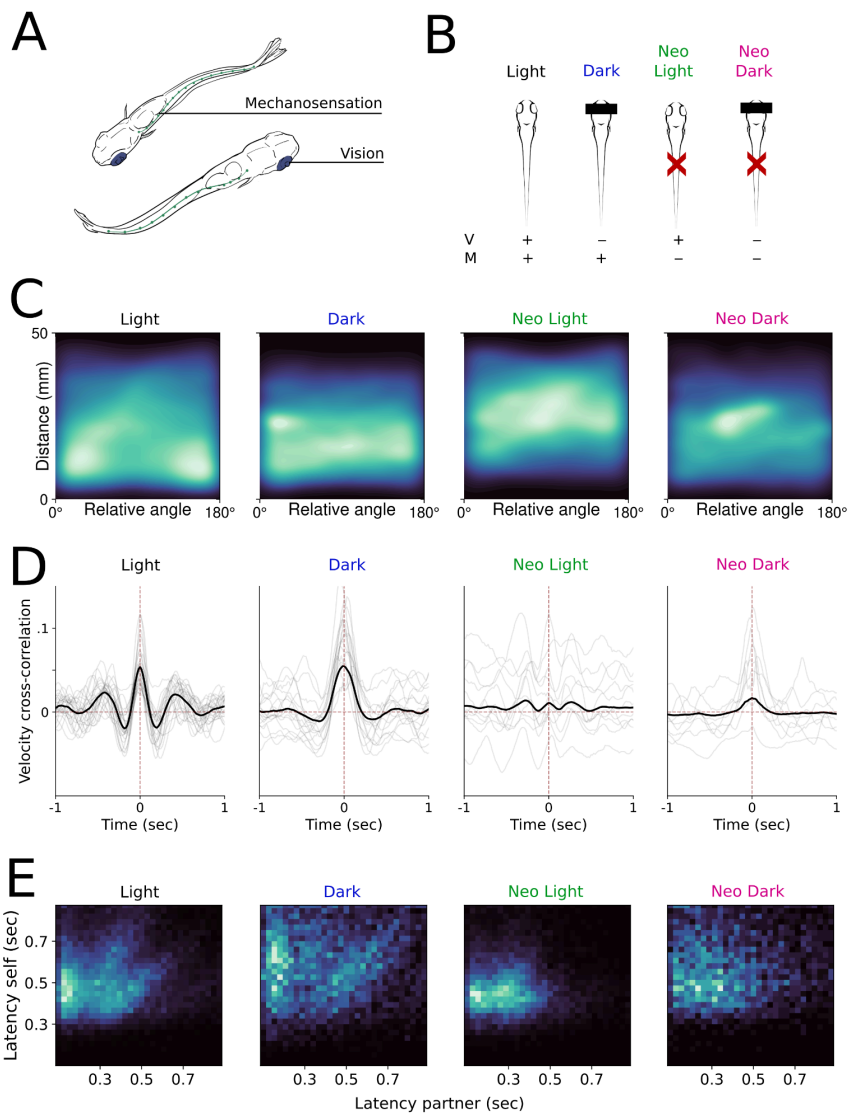

**Figure 5. A.)** Illustration of the sensory systems manipulated in the following experiments. **B.)** Schematic of the sensory manipulations for all four groups. **C.)** 2D histograms of distance and relative angle for all conditions. **D.)** Time-lag velocity cross-correlations for pairs in different sensory conditions. **E.)** Relative interbout interval coordination for all conditions represented by phase response curves.

We then used our cLM-HMM to classify social interactions under these four conditions, with the goal of revealing which sensory modalities are important for the states derived from our model. Across each condition, pairs spend different proportions of time in each state, suggesting that states relate to sensory-specific behaviors. We evaluated this observation statistically using vision and mechanosensation as categorical variables in a mixed model with the pair of origin as a random effect (Fig 6A-B). Synchronized in-phase interactions decrease when neuromasts are ablated, but are still present in the dark, indicating that water flow contributes importantly to this mode of interaction, though we did not find this effect to be statistically significant (p = 0.055, Fig 6B). Conversely, delayed out-of-phase interactions are reduced in the dark but enhanced in neo-light conditions even relative to the light condition, suggesting that fish rely on visual cues for the maintenance of mid-range alignment in the absence of mechanosensation (p < 0.001, Fig 6B). Vision primarily determines whether or not pairs inhabit the null state (p = 0.001), and we found no interaction effects between vision and mechanosensation for any state, suggesting that vision is the dominant sensory modality. Importantly, our model effectively captures that movements become uncoordinated in all conditions without the full complement of sensory information and that such pairs dwell in the null state for greater durations, suggesting that access to both sensory modalities optimally supports sustained interactions (Fig 6C).

Similarly, pairs in the dark sustain synchronized interactions for longer durations, and the duration of these interactions is distance-dependent only in pairs with intact mechanosensation (Fig 6C, S3A). Further, partner fish are more likely to be in the field of view during delayed interactions, and this is particularly pronounced in neo-light pairs (Fig S2D). This finding supports our prediction that positioning the partner outside of the blind spot is both visually mediated and occurs as a result of social interaction, and further suggests that the delayed interaction state in our model corresponds to visually-driven behaviors. The sensory-specific nature of each state is further supported by our observation that pairs increase the proportion of time spent in the state that corresponds to their available sensory information, relative to pairs with all sensory inputs, which alternate between the two states.

We characterized the distance and relative heading of pairs segmented by state across the four conditions, finding that synchronized interactions tend to occur at close distances at either parallel or antiparallel alignments, while these effects are less pronounced in the delayed state (Fig 6D). Interestingly, this alignment is also observed in the null state in control animals, likely a result of brief transitions to the null state before re-engaging. Transition probabilities indicate that pairs in the dark are most likely to transition to the null state from interacting states, while pairs in the light alternate between the two interaction states (Fig 6E). We further analyzed the configurations at the onset of a given interaction state, finding that under normal conditions, state transitions occur primarily at parallel and antiparallel alignments (Fig 6F), while other conditions show variable distances and alignments. This spatial arrangement suggests that pairs switch between interaction states at the stereotyped orientations we previously identified, such that synchronized states begin with parallel alignment, null states begin when anti-aligned, and both configurations represented at the onset of delayed interactions.

Although the cLM-HMM fit to the whole dataset yields a single set of interaction weights $^{(k)}$ shared by all fish pairs, we can investigate whether the signatures of the three interaction states were modulated across different pairs and, in particular, across the four conditions. We thus estimated the "local weights" for each state $k$, but conditioned only on observations from a single fish pair that experiences one condition (see Methods). We found that local weights fit to individual pairs show considerable variability under normal conditions, and this raised the possibility that weights would differ across sensory conditions (Fig S3D). By further averaging local weights for all pairs within a particular sensory condition, we found that synchronized states in dark pairs exhibit just a single peak at short lag, lacking the quasi-periodic feature observed during the light and neo-light condition (Fig S3C). This suggests that mechanosensation drives only near-instantaneous feedback, consistent with the lateral line's role in other time-critical behaviors like predator avoidance (Stewart et al, 2013). In contrast, delayed components are evident in neo-light pairs even in the synchronized state. High variability across pairs was found for local weights in the delayed state in both dark conditions, as well as null states in both light conditions, reflecting the very low occupancy for those states in the respective conditions.

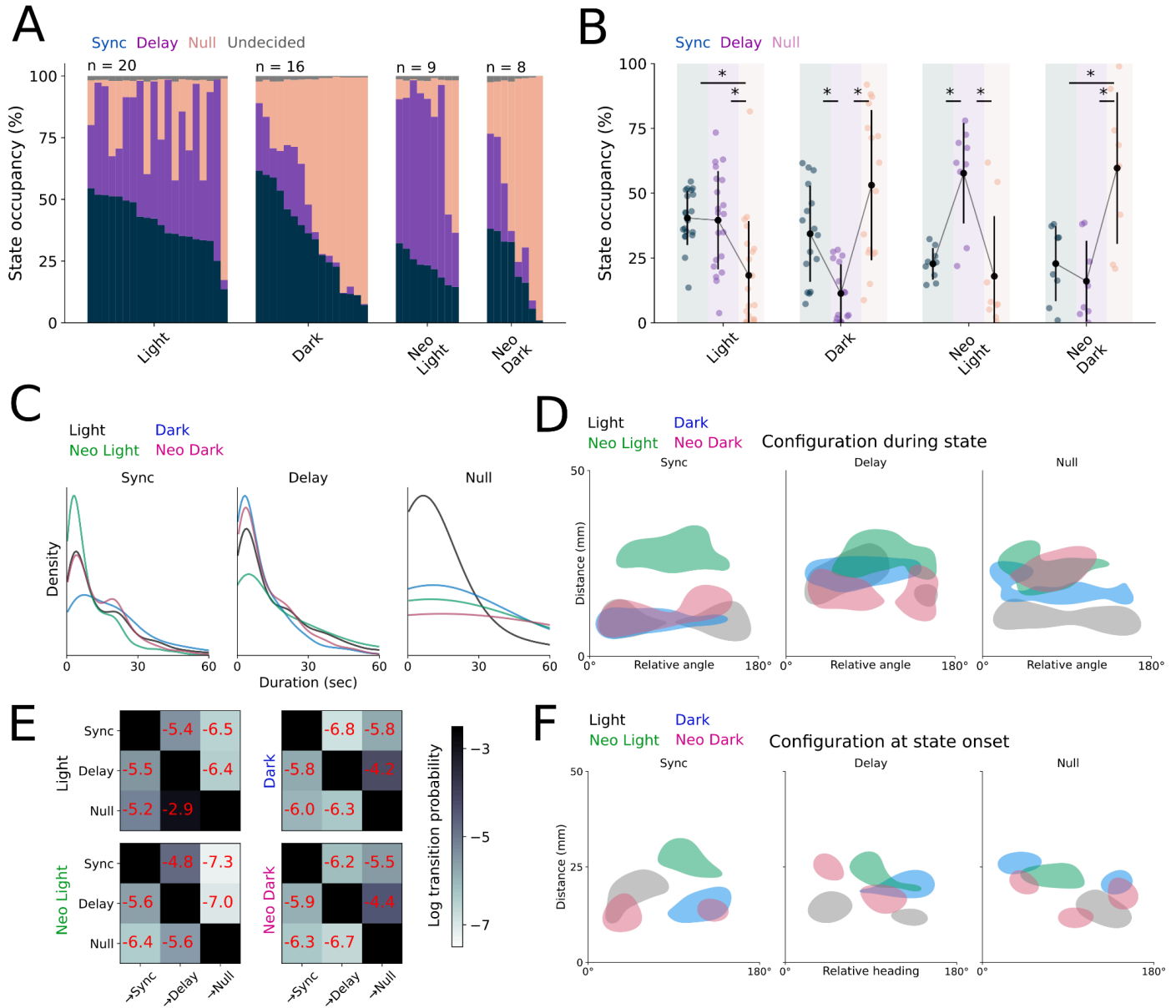

**Figure 6. A.)** State occupancy as classified by our cLM-HMM for all pairs across sensory conditions. Each column represents one pair. **B.)** State occupancy as classified by our cLM-HMM, excluding the undecided state. Each trio of data points represents one pair. **C.)** Kernel density estimate for state durations across sensory conditions. **D.)** 2D histograms of relative angle and distance for all sensory conditions conditioned by state. Shaded areas represent the 20% highest densities of the distribution. **E.)** Non-self lexical transition probabilities between states for all sensory conditions in log scale. **F.)** 2D histograms of relative angle and distance for all sensory conditions at the initial transition to a given state. Shaded areas represent the 20% highest densities of the distribution. Asterisks indicate a p value below an alpha of 0.05 corrected for multiple comparisons with a Šidák adjustment.

Altogether, we observed that visual input results in slower time scale interactions at greater distances and is important for maintaining the position of other animals in the field of view. Conversely, mechanosensory input is important for fast, coordinated swimming while parallel. Proximity, and therefore the strength of local mechanosensory input, is important for both initiating and maintaining fast interactions. Animals deprived of both forms of sensory feedback show a strong suppression of social interaction as measured in our assay, suggesting that visual and mechanosensory input are the dominant sensory cues in the behaviors assayed here.

## DISCUSSION

We combined experimental and theoretical approaches to develop a novel model that identified unique states of coordinated interactions in fish that alternate over time. By applying our model to experiments under different combinations of sensory conditions, we revealed that both mechanosensory and visual input are necessary for normal social behavior. Additionally, we found that these two sensory modalities operate on distinct time scales and support specific spatial features exhibited by fish during social behavior. Interestingly, pairs of fish adjust their interactions to take advantage of the sensory modality available to them, compensating partially for the loss of sensory information. Precisely how and why animals under normal conditions reach a decision boundary to switch between these modalities is an open question.

Our modeling approach is based on the idea that social interactions require the coordination of movements between paired animals, allowing us to predict the movements of one fish from the movements of its partner. Moreover, we showed that the predictive relationships between the two fish, as encoded in the model weights, are not stationary. Instead, they vary on a moment-to-moment basis and can be captured by a Markov chain. The resulting hidden Markov model with constrained linear-model emissions represents a new and parsimonious way to uncover the spatiotemporal features of social behavior with wide-ranging applications to other pairwise interactions.

An HMM-based approach with categorical emissions was previously used to predict which song a male fly would deploy during mating behavior (Calhoun et al, 2019). These experiments used the joint probability of movements for both animals to identify distinct states that govern song choice, demonstrating that the sensorimotor filters animals use can change depending on context, and highlighting the importance of internal states in behavioral choice. Recent work also showed that joint probabilities of kinematics of two animals can be used to identify different features of fighting behavior in adult zebrafish (O'Shaughnessy et al 2023). In our data, we found that models fit on the joint probabilities of movements of both fish failed to yield states significantly different from shuffled data, likely due to the strong autocorrelation of each fish time series (not shown; see below). A central tool in our investigation is the assessment of the validity of our analyses by comparing empirical data with surrogate data obtained by shuffling fish between different pairs to control for nonsense correlations between fish movements. The presence of strong autocorrelations in time series, such as the inherent periodicity of single-fish bouts, can lead to mirage or nonsense correlations (Harris 2021, Yule 1926).

This practical limitation of our data required us to use an approach that differed in several respects. First, we predicted movements of one animal from the other animal, rather than using the joint probability of movements for both animals. With this method, the Linear Model weights we derived were inherently based on the temporal relationship between the behavior of both animals, suppressing the influence of irrelevant autocorrelations. We introduced constraints on the weights in each state, based on symmetric Toeplitz matrices, which vastly reduces the number of model parameters to fit, leading to efficient and robust inference of the interaction states from a limited number of samples. Crucially, we found that the cLM-HMM fit to the shuffled data did not exhibit any meaningful weights, thus corroborating our results on the empirical data.

Constrained LM-HMMs are a new class of models that are advantageous in experimental situations with limited sample size, including naturalistic behaviors such as social interactions. In this case, reducing the number of model parameters by introducing constraints on the model emission distributions will be crucial to obtain efficient and robust inference with limited sample size. In combination with the efficiency gains from reducing the number of model parameters, this allowed us to obtain unique "local" weights for single fish pairs by applying the M-step of the EM algorithm on the observations restricted to a single pair, given the posterior probabilities from the global fit. This approach both reduces error in assigning states across different experiments, and captures more of the underlying variability by revealing subtle differences across individuals and sensory modalities.

Although we found only three states in juveniles, we anticipate that social behavior in later developmental stages or in different species will exhibit a larger repertoire of interaction modes that occur sparsely, each of which can be represented by an HMM hidden state. One example is cooperative biparental care that requires task-specific investment in offspring from two individuals, and is described in birds, mammals, fishes, and even insects such as *Nicrophorus* burying beetles (Benowitz & Moore 2016, Cockburn 2006, Gromov 2011, Gross & Sargent 1985). Some emergent collective behaviors in animals considered behaviorally advanced can be partially explained by simple rules modified by relationships between individuals, as in the case of baboons during troop movements (Farine et al 2016, King et al 2011). However, incorporating subtle sensory cues into these models could reveal greater underlying complexity, and comparatively brief behavioral states between individuals that initiate broader collective transitions could be represented in a probabilistic model like the one presented here.

Moreover, we revealed that each interaction state is encoded in a particular linear model, and that these states differ by the time lags of coordination. This unique approach allowed us to identify the temporal dynamics and distinct biologically relevant features of a complex behavior starting from simple time series data of the fish speed. A version of our model that only relies on past information is feasible and would permit the inference of interacting states in real time, enabling closed loop experiments that manipulate sensory information or brain activity only during the execution of specific behaviors. This approach could enable the real-time mapping of behavioral circuits and their contributions to dynamic social behaviors.

To make our experiments tractable, we observed only two fish in a depth-constrained enclosure. However, in their natural habitat, zebrafish occupy diverse environments with unique spatial properties, and form groups of many individuals. With three states defined for pair-wise interactions, it will be interesting to explore whether these dynamics extend to larger areas or groups. Given that local interactions necessarily govern the collective behavior of large groups of animals due to the limited sensory information available to an individual, such models of local interactions have the potential to recapitulate broader group dynamics (Couzin et al 2002, Davidson et al 2021, Heras et al 2019, Lemasson et al 2009). In particular, the coupled oscillatory system we describe here is broadly applicable across both biological and physical systems, and our model could be used to perform unsupervised classification of collective states in any system based on correlations between individual units (O'Keeffe et al 2017, Reidl et al 2023, Winfree 1967).

Addressing these limitations and potential confounds in future work might allow us to improve the characterization of interaction states and potentially identify more sensory-specific states beyond the synchronized and delayed ones identified here. Although we identified biologically plausible sensory mechanisms by using a relatively simple time series of individual velocities, new models could also incorporate the specific sensory information experienced by each animal (Davidson et al 2021, Harpaz et al 2021, Lemasson et al 2009), rather than measuring them as an output of our state classification. Further, the neural circuits underlying these behaviors are incompletely described, though many evolutionarily homologous brain regions are implicated in the teleost social decision making network, including the thalamus, tectum, and telencephalon (Anneser et al., 2020, Groneberg et al., 2020, Kappel et al 2022, Pinho et al 2023, Shinozuka & Watanabe 2007, Stednitz et al 2018, Wee et al, 2022). Targeted anatomical manipulations of regions in conjunction with our behavioral classification could reveal the anatomical circuits required for specific aspects of these behaviors. A recording platform that enables brain-wide or circuit-level calcium imaging in pairs of freely behaving zebrafish is an exciting future direction for disentangling the temporal relationship between stimulus and behavioral response.

# METHODS

## Fish husbandry

TU *mitfa* -/- zebrafish were maintained under standard housing conditions at 28°C on a 14 hour light cycle as described in (Westerfield, 2007) at the University of Oregon, University of Queensland, and University of Melbourne. Juvenile zebrafish were reared on a diet of dry feed supplemented with rotifers. Experiments were performed in accordance with approval 2021/AE001047 from the University of Queensland Animal Welfare Unit and approval 2022-24987-35220-5 from the University of Melbourne Office of Research Ethics and Integrity.

## Behavioral recordings & analysis

Juvenile experiments took place in a custom-built light proof cabinet. Zebrafish were placed as pairs into behavioral chambers constructed of 1mm thick clear PDMS with a 50 mm circular arena cut into the center, bonded to glass slides. All pairs comprised siblings reared in the same tank to minimize confounds from novel social partners. The tops of the chambers were sealed using a glass coverslip to ensure a consistent fluid volume in the arena. Illumination was achieved with a ring of infrared LEDs (SMD5050-300-IR 850 nm), and imaged with a FLIR Blackfly S-USB3 camera (BFS-u3-122S6M-C) for 15 minutes. A white LED was used to illuminate the arena in light condition experiments, and this LED was switched off for dark condition experiments. For the experiments in Fig 1-4, recordings were performed at 60 fps and a resolution of 18.9 pixels/mm.

Tracking from videos was achieved using SLEAP (Pereira et al 2022) to locate the head and body. These positions were used to calculate all subsequent variables. Further behavioral analysis was performed using custom software written in Julia 1.8.5.

Adult (3 month old) zebrafish experiments (Supplemental Fig S1B, S2A) were performed in a 25 cm acrylic arena with 10cm water depth, recorded at 10 fps with a Mightex B3-U12 camera under infrared light and analyzed using custom software written in Python (github.com/stednitzs/daniopen/).

## Chemical ablation of neuromasts

Neuromasts of the lateral line were ablated by bath immersion in 200 uM neomycin prepared in standard embryo medium for one hour. Animals were allowed to recover in fresh embryo medium (juveniles) or aquarium water (adults) for 2 hours prior to behavioral recording. The efficacy of this protocol was first confirmed with DASPEI staining to fluorescently label neuromasts.

## Modeling

The fish speeds were obtained by calculating the distance traversed by SLEAP tracked head points between frames. Recordings of each pair were filmed at 60 FPS for 15 minutes with a resolution of 18.9 pixels/mm. Given the discrete burst-glide movement pattern of larval zebrafish, we converted fish speeds into binary bout sequences where 1 or 0 indicates an active or inactive bout, respectively. We determined a bout is activated when a threshold speed of 5.4 mm/s is surpassed for the next 67 ms and terminated when the speed falls below 3.8 mm/s for the next 50 ms.

To test the hypothesis that fish movement between pairs is partner-driven and characterized by unique cross-correlated states, we develop a HMM with input-driven spherical-Gaussian observations modeling the time-varying linear dependence

$$\vec{y}_t = W^{(k)}\vec{x}_t + b^{(k)}\vec{1} + \vec{\epsilon}_t$$

between the symmetrically time-lagged speeds of the focal fish $\vec{y}_t$ with the input fish $\vec{x}_t$ for states $k = 1,..., K$. We denote the binary bout sequences (interchangeably referred to as speed) of the input and focal fish as $v_1(n)$ and $v_2(n)$, respectively, for $n = 1,..., T + 2L$ and construct the $D = 2L + 1$ dimensional input and output vectors as

$$\vec{x}_t = [v_1(t), ..., v_1(t+L), ..., v_1(t+2L)]^\top$$
$$\vec{y}_t = [v_2(t), ..., v_2(t+L), ..., v_2(t+2L)]^\top$$

for $t = 1,..., T$. Each vector symmetrically lags the respective fish speed $L$ steps about index $t + L$.

We model the bias $b^{(k)}$ as a constant offset and $\epsilon_t$ as i.i.d Gaussian noise with zero mean and spherical covariance $\Sigma^{(k)} = \sigma^2 I$ tied across all states. We constrain the $D \times D$ weights matrix $W^{(k)}$ to have a symmetric Toeplitz structure. Other model parameters include the transition matrix $A$, which captures the probability $A_{kl} = p(z_{t+1} = k | z_t = l)$ of switching from state $l$ at time $t$ to state $k$ at time $t+1$, and the initial state probabilities $\pi^{(k)}$. We use the Expectation–Maximization (EM) algorithm to maximize the expected complete-data log likelihood (ECLL) and obtain the optimized parameters $\Theta = \{W^{(k)}, b^{(k)}, \sigma^2, A, \pi^{(k)}\}$. We refer to this model as a constrained linear-model HMM (cLM-HMM).

To gain insight into the function of the weights matrix, and to understand how the Toeplitz structure arises, we first consider maximization of the unconstrained matrix. The contribution to the ECLL from our observation model is:

$$Q(W^{(k)}, b^{(k)}, \sigma^2) = -\frac{1}{2}\sum_{t=1}^{T}\sum_{k=1}^{K}\gamma(z_{tk})[D\ln\sigma^2 + \frac{1}{\sigma^2}||\vec{y}_t - W^{(k)}\vec{x}_t - b^{(k)}\vec{1}||^2]$$

where $\gamma(z_{tk})$ is efficiently calculated during the E-step of the EM algorithm and denotes the posterior probability of being in state $k$ at time $t$ given the data and current model parameters (Bishop & Nasrabadi 2006). Joint maximization of the ECLL with respect to $b^{(k)}$ and $W^{(k)}$ yields a system of equations for each row $\vec{w}_i^{(k)}$ of the weights matrix

$$R_{21}^{(k)}(j-i) \approx \frac{\sigma_2^{(k)}}{\sigma_1^{(k)}}[\vec{w}_i^{(k)} \star R_1^{(k)}](j)$$

for indices $i, j = 0,..., 2L$ where we define $R_{pq}^{(k)}(\tau)$ as the state-conditioned cross-correlation ($p \neq q$) or autocorrelation ($p = q$) function of the respective fish speeds corresponding to indices $p, q = 1, 2$. Similarly, we define $\sigma_p^{(k)}$ as the state-conditioned standard deviation of $v_p(t)$. In general terms, each row $\vec{w}_i^{(k)}$ serves as a filter that identifies unique states of cross-correlated structure between the output and input time series when convolved with the autocorrelation of the input series, scaled by the ratio of state-conditioned standard deviations of the respective signals. The shift invariance of this expression with respect to indices $i$ and $j$ implies that each row $\vec{w}_{i+1}^{(k)}$ must be a one-step time-shifted version of the preceding row $\vec{w}_i^{(k)}$, resulting in a weights matrix with diagonal-constant (Toeplitz) structure. This expression holds for $T \gg L$ given the absence of significant outliers in the first and last $2L$ steps of both the input and output time series.

Although symmetry of the weights matrix does not naturally emerge from the expression (since cross-correlation is not inherently symmetric), the process of fitting our model to numerous pairs while considering both permutations of the input and focal fish effectively averages out asymmetries, leading to symmetric weights. We therefore constrain the weights matrix to have symmetric Toeplitz structure, reducing the number of free parameters from $D^2$ to $D$. As a result of this constraint, the M-step cannot be performed in closed form and must be carried out using a numerical optimization algorithm. To reduce model complexity, we downsample each input and output vector by a factor of $q = 5$ after construction.

To ensure our model identifies genuine interaction states rather than noise, we compare fits on empirical data to those on shuffled data obtained from splicing together randomly drawn input and focal fish from different pairs. Theoretically, since each row of the weights matrix is a one-step time-shifted version of the previous row, identifying the central row $\vec{w}_L^{(k)}$ is sufficient to capture the bulk of the filter. This can be achieved by lagging the output vector as usual while constructing the one-dimensional input vector $\vec{x}_t = [v_1(t + L)]$. In practice, we find that states identified in the empirical data with this approach are indistinguishable from those found in the shuffled data due to the transient "nonsense correlations" that emerge by chance in strongly periodic signals (Harris 2021, Yule 1926). We find that our model captures genuine, sustained interactions when the lag $L$ for both the input and output vectors is longer than the autocorrelation time of the respective signals but shorter than the interaction duration between animals. For our system, a lag of $L = 2$ seconds is sufficient.

In total, we collected data from $N = 53$ pairs of juvenile zebrafish aged 12-13 dpf across varying sensory modalities, including light and dark environments with either intact or ablated hair cells. Specifically, in the light environment, hair cells were intact for 20 pairs and ablated for 9. In the dark environment, hair cells were intact for 16 pairs and ablated for 8.

To identify the observation model parameters $\phi^{(k)} = \{W^{(k)}, b^{(k)}, \sigma^2\}$ that describe the global set of interaction states available to the animals, we fit multiple instances of randomly initialized models to all dyads across all conditions and permutations of input and focal fish. We determined that three states were optimal through a combination of cross-validation and the observation that additional states beyond three appeared as slight variations of previously existing states. We selected the model with the highest likelihood for $K = 3$ states which correspond to in-phase (synchronized), out-of-phase (delayed), and non-interacting motion. We assign state $k$ to time point $t$ by thresholding the argmax of the posterior probability $\gamma(z_{tk}) > 0.8$. If the threshold is not surpassed, we denote the time point as undecided.

To test model consistency, we compared state sequences between permutations of the selected input and focal fish within each pair. We identified semi-frequent mismatches in state assignments between the delay and null states when swapping the input and focal fish, likely due to the relatively large standard deviation of the model ($\sigma = 0.34$) for signals ranging between 0 and 1. To ensure our model assigns the state that best describes pairwise interactions regardless of input and focal fish selection, we construct a dual-observation model that simultaneously fits both permutations within each pair. We hold fixed the observation model parameters $\phi^{(k)}$ obtained previously from global fits on our data and run only the E-step of the EM algorithm individually for each pair. This allows us to obtain local posteriors and thereby state sequences that are tied across both permutations of the selected input and focal fish.

Finally, we fix the local posteriors obtained from the dual-observation model and run the single-observation model M-step individually for each pair and permutation of input and focal fish. This approach yields the local model parameters $\phi_p^{(k)} = \{W_p^{(k)}, b_p^{(k)}, \sigma_p\}$ for each pair and permutation $p$, enabling us to uncover interaction state variability between pairs and across different sensory conditions.

## ACKNOWLEDGMENTS

The authors would like to thank the staff of aquatic services at the University of Oregon, University of Queensland, and the University of Melbourne for assistance with zebrafish colony maintenance and rearing of juveniles. We would also like to thank Johannes Larsch for invaluable comments and discussions regarding the project. L.M. was partially supported by National Institutes of Health grants R01NS118461, R01MH127375 and R01DA055439 and National Science Foundation CAREER Award 2238247. A.L. was supported by the National Science Foundation Graduate Research Fellowship under Grant No. 2236419. E.K.S. was supported by a Simons Foundation Research Award (625793), two ARC Discovery Project Grants (DP220103812 & DP230102614), and an NHMRC Investigator Award (2027072). The research reported in this publication was supported by the National Institute of Neurological Disorders and Stroke of the National Institutes of Health under Award Number R01NS118406 to E.K.S. The content is solely the responsibility of the authors and does not necessarily represent the official views of the National Institutes of Health. Any opinion, findings, and conclusions or recommendations expressed in this material are those of the authors(s) and do not necessarily reflect the views of the National Science Foundation.

## AUTHOR CONTRIBUTIONS

S.J.S, E.K.S., and P.W. designed the project; S.J.S. performed experiments with juvenile fish; A.L.F. and S.J.S. performed experiments with adult fish; A.L. and L.M. designed and fit the model; A.L. and S.J.S. analyzed the data, with help from L.M. and P.P.; A.L. and S.J.S. produced the figures; S.J.S., A.L., L.M., P.W., E.K.S. wrote the manuscript; E.K.S., P.W. and L.M. supervised the project.

## CITATIONS

# SUPPLEMENTARY MATERIAL

## Relationship between weights, auto and cross-correlation:

Following the notation in the Methods section, let $v_1(t)$ and $v_2(t)$ denote the speeds of the input and focal fish, respectively, for $t = 1,...,T$. We hypothesize that the speed of the focal fish can be predicted from that of the input fish

$$v_2(t) = [W \star v_1](t) + b$$

where the $\star$ operator denotes the discrete convolution between a filter $W$ of length $D$ and $v_1$

$$[W \star v_1](t) = \sum_{l=0}^{D} W(l) v_1(t - l) + b.$$

Next, we define the normalized discrete cross-correlation function between signals $v_p(t)$ and $v_q(t)$ for $p, q = 1, 2$

$$R_{pq}(\tau) \equiv \frac{1}{\sigma_p \sigma_q T} \sum_{t=1}^{T} \left[v_p(t) - \mu_p\right]\left[v_q(t+\tau) - \mu_q\right] = \frac{1}{\sigma_p \sigma_q T} \sum_{t=1}^{T} \left[v_p(t) - \mu_p\right] v_q(t+\tau)$$

where we expand and use the definition of the mean $\mu_p \equiv \frac{1}{T}\sum_{t=1}^{T} v_p(t)$ to arrive at the final equality. If $p = q$, then we define $R_p(\tau) \equiv R_{pp}(\tau)$ as the autocorrelation function. We let $v_q(t+\tau) = 0$ if index $t + \tau$ is outside the range of $v_q$.

Inserting the expression for $v_2(t+\tau)$ into the cross-correlation function, we find

$$R_{21}(\tau) = \frac{\sigma_1}{\sigma_2} \sum_{l=0}^{D} W(l) R_1(l - \tau) = \frac{\sigma_1}{\sigma_2} [W * R_1](\tau)$$

demonstrating that unique filters $W$ can be identified that relate state-conditioned autocorrelated movement from some input signal to the cross-correlation with the focal signal.

## M-step update equations for unconstrained weights, constant biases:

We now consider multiple linear models for states $k = 1,...,K$ using a hidden Markov model (HMM) framework. Following Bishop & Nasrabadi 2006, we optimize the expected complete-data log likelihood (ECLL)

$$Q(\Theta, \Theta^{(t-1)}) = \sum_{k=1}^{K} \gamma(z_{1k}) \ln \pi^{(k)} + \sum_{t=2}^{T} \sum_{j=1}^{K} \sum_{k=1}^{K} \xi(z_{t-1,j}, z_{tk}) \ln A_{jk} + \sum_{t=1}^{T} \sum_{j=1}^{K} \gamma(z_{tk}) \ln p(\vec{y}_t | \vec{x}_t, \phi^{(k)})$$

where the posteriors $\gamma(z_{tk})$ and joint posteriors $\xi(z_{t-1,j}, z_{tk})$ are efficiently calculated in the E-step of the Expectation-Maximization (EM) algorithm. The observation model is defined by $p(\vec{y}_t | \vec{x}_t, \phi^{(k)})$ which gives the probability of observing the output $\vec{y}_t$ given the input $\vec{x}_t$ and observation model parameters $\phi^{(k)}$.

We denote the speed of the input and focal fish as $v_1(n)$ and $v_2(n)$, respectively, for $n = 1,...,T + 2L$ and construct the $D = 2L + 1$ dimensional input and output vectors

$$\vec{x}_t = [v_1(t), ..., v_1(t+L), ..., v_1(t+2L)]^\top$$

$$\vec{y}_t = [v_2(t), \ldots, v_2(t+L), \ldots, v_2(t+2L)]^T$$

for $t = 1,\ldots, T$. Each vector captures the symmetrically time-lagged speeds about index $t + L$.

We consider an input-driven spherical-Gaussian observation model with covariance $\Sigma^{(k)} = \sigma^2 I$ tied across states

$$p(\vec{y}_t | \vec{x}_t, W^{(k)}, \vec{b}^{(k)}, \sigma^2) = N(\vec{y}_t | W^{(k)} \vec{x}_t + \vec{b}^{(k)}, \sigma^2 I)$$

The contribution of the observation model to the ECCL is

$$Q(W^{(k)}, \vec{b}^{(k)}, \sigma) = -\frac{1}{2} \sum_{t=1}^{T} \sum_{k=1}^{K} \gamma(z_{tk}) [D \ln 2\pi + D \ln \sigma^2 + \frac{1}{\sigma^2} || \vec{y}_t - W^{(k)} \vec{x}_t - \vec{b}^{(k)} ||^2]$$

For applications within this paper, we constrain the biases to be a constant vector $\vec{b}^{(k)} = b^{(k)} \vec{1}$ and the $D \times D$ weights matrix $W^{(k)}$ to have a symmetric Toeplitz structure. The constraints placed on $W^{(k)}$ require the M-step be carried out using a numerical optimization algorithm. We refer to this model as a constrained linear-model HMM (cLM-HMM).

To gain insight into the role of the weights matrix, and to see how the Toeplitz structure naturally arises, we consider maximization of the unconstrained model parameters. Maximization of the ECLL with respect to the biases reveals

$$\vec{b}^{(k)} = \frac{\sum_{t=1}^{T} \gamma(z_{tk}) [\vec{y}_t - W^{(k)} \vec{x}_t]}{\sum_{t=1}^{T} \gamma(z_{tk})} \equiv \vec{\mu}_2^{(k)} - W^{(k)} \vec{\mu}_1^{(k)} \approx \mu_2^{(k)} \vec{1} - \mu_1^{(k)} W^{(k)} \vec{1}$$

where we define the $i^{th}$ element of the state-conditioned mean vector $[\vec{\mu}_p^{(k)}]_i = \mu_p^{(k)}(i) \equiv \frac{\sum_{t=1}^{T} \gamma(z_{tk}) v_p(t+i)}{\sum_{t=1}^{T} \gamma(z_{tk})}$ for $i = 0,\ldots, 2L$ and $p = 1, 2$. We note $\mu_p^{(k)}(i) \approx \mu_p^{(k)}(j)$ for all $i \neq j$ when $T \gg L$ barring any significant outliers within the first and last $2L$ steps. We refer to this criteria as the short-lag regime. For simplicity, we define the state-conditioned mean $\mu_p^{(k)} \equiv \mu_p^{(k)}(L)$ and approximate $\vec{\mu}_p^{(k)} \approx \mu_p^{(k)} \vec{1}$. Similarly, we define the state-conditioned standard deviation $\sigma_p^{(k)}(i) = \left( \frac{\sum_{t=1}^{T} \gamma(z_{tk}) [v_p(t+i) - \mu_p^{(k)}(i)]}{\sum_{t=1}^{T} \gamma(z_{tk})} \right)^{1/2}$ and let $\sigma_p^{(k)} \equiv \sigma_p^{(k)}(L)$.

Maximization of the ECLL with respect to the unconstrained weights matrix results in solving $W^{(k)} B^{(k)} = C^{(k)}$ where

$$B^{(k)} \equiv \sum_{t=1}^{T} \gamma(z_{tk}) [\vec{x}_t - \vec{\mu}_1^{(k)}] \vec{x}_t^T$$

$$C^{(k)} \equiv \sum_{t=1}^{T} \gamma(z_{tk}) [\vec{y}_t - \vec{\mu}_2^{(k)}] \vec{x}_t^T$$

Each entry $i, j = 0,\ldots, 2L$ of the respective matrices can be expressed as

$$B_{ij}^{(k)} \approx \sum_{t=1}^{T} \gamma(z_{tk}) [v_1(t+i) - \mu_1^{(k)}] v_1(t+j) = \sum_{t=1+i-L}^{T+i-L} \gamma(z_{t-i+L, k}) [v_1(t+L) - \mu_1^{(k)}] v_1(t+L+j-i)$$

$$C_{ij}^{(k)} \approx \sum_{t=1}^{T} \gamma(z_{tk}) [v_2(t+i) - \mu_2^{(k)}] v_1(t+j) = \sum_{t=1+i-L}^{T+i-L} \gamma(z_{t-i+L, k}) [v_2(t+L) - \mu_2^{(k)}] v_1(t+L+j-i)$$

where we approximated $\vec{\mu}_p^{(k)} \approx \mu_p^{(k)} \vec{1}$ to arrive at the first equality and re-indexed in the last equality. Next, we define the normalized, state-conditioned cross-correlation between signals $v_p$ and $v_q$

$$R_{pq}^{(k)}(\tau) \equiv \frac{1}{\beta_{pq}^{(k)}} \sum_{t=1}^{T} \gamma(z_{tk})[v_p(t+L) - \mu_p^{(k)}]v_q(t+L+\tau)$$

where the normalization constant $\beta_{pq}^{(k)} \equiv \sigma_p^{(k)} \sigma_q^{(k)} \sum_{t=1}^{T} \gamma(z_{tk})$. If $p = q$, we define $R_p^{(k)}(\tau) \equiv R_{pp}^{(k)}(\tau)$ as the autocorrelation function. Using this expression, we can approximate

$$B_{ij}^{(k)} \approx \beta_1^{(k)} R_1^{(k)}(j-i)$$
$$C_{ij}^{(k)} \approx \beta_{21}^{(k)} R_{21}^{(k)}(j-i)$$

which is valid in the short-lag regime when the duration of each state is typically longer than the lag $L$. We can further express $C_{ij}^{(k)} = \sum_{l=0}^{2L} W_{il}^{(k)} B_{lj}^{(k)}$. Combining expressions, we find

$$R_{21}^{(k)}(j-i) \approx \frac{\sigma_1^{(k)}}{\sigma_2^{(k)}} [\vec{w}_i^{(k)} * R_1^{(k)}](j)$$

where $\vec{w}_i^{(k)}$ denotes the $i^{th}$ row of the weights matrix. The shift invariance of this expression with respect to indices $i$ and $j$ implies that each row $\vec{w}_{i+1}^{(k)}$ must be a one-step time-shifted version of the preceding row $\vec{w}_i^{(k)}$. This results in a weights matrix with approximately Toeplitz structure.

### Dual Fit Model

To simultaneously fit both permutations of the selected input and output signals, we construct the column-stacked $D = 4L + 2$ dimensional dual input $\vec{x}_t^{(2)} = \begin{bmatrix} \vec{x}_t^T & \vec{y}_t^T \end{bmatrix}^T$ and output $\vec{y}_t^{(2)} = \begin{bmatrix} \vec{y}_t^T & \vec{x}_t^T \end{bmatrix}^T$ vectors where the single input $\vec{x}_t$ and output $\vec{y}_t$ vectors are as previously defined. We define the dual-observation model

$$p(\vec{y}_t^{(2)} | \vec{x}_t^{(2)}, \phi^{(k)}) = p(\vec{y}_t | \vec{x}_t, \phi^{(k)}) p(\vec{x}_t | \vec{y}_t, \phi^{(k)}) = N(\vec{y}_t | W^{(k)} \vec{x}_t + b^{(k)} \vec{1}, \sigma^2 I) N(\vec{x}_t | W^{(k)} \vec{y}_t + b^{(k)} \vec{1}, \sigma^2 I)$$

resulting in the contribution to the ECLL

$$Q(W^{(k)}, b^{(k)}, \sigma) = -\frac{1}{2} \sum_{t=1}^{T} \sum_{k=1}^{K} \gamma(z_{tk})[ 2D \ln 2\pi + 2D \ln \sigma^2 + \frac{1}{\sigma^2} || \vec{y}_t - W^{(k)} \vec{x}_t - b^{(k)} \vec{1} ||^2 + \frac{1}{\sigma^2} || \vec{x}_t - W^{(k)} \vec{y}_t - b^{(k)} \vec{1} ||^2 ]$$

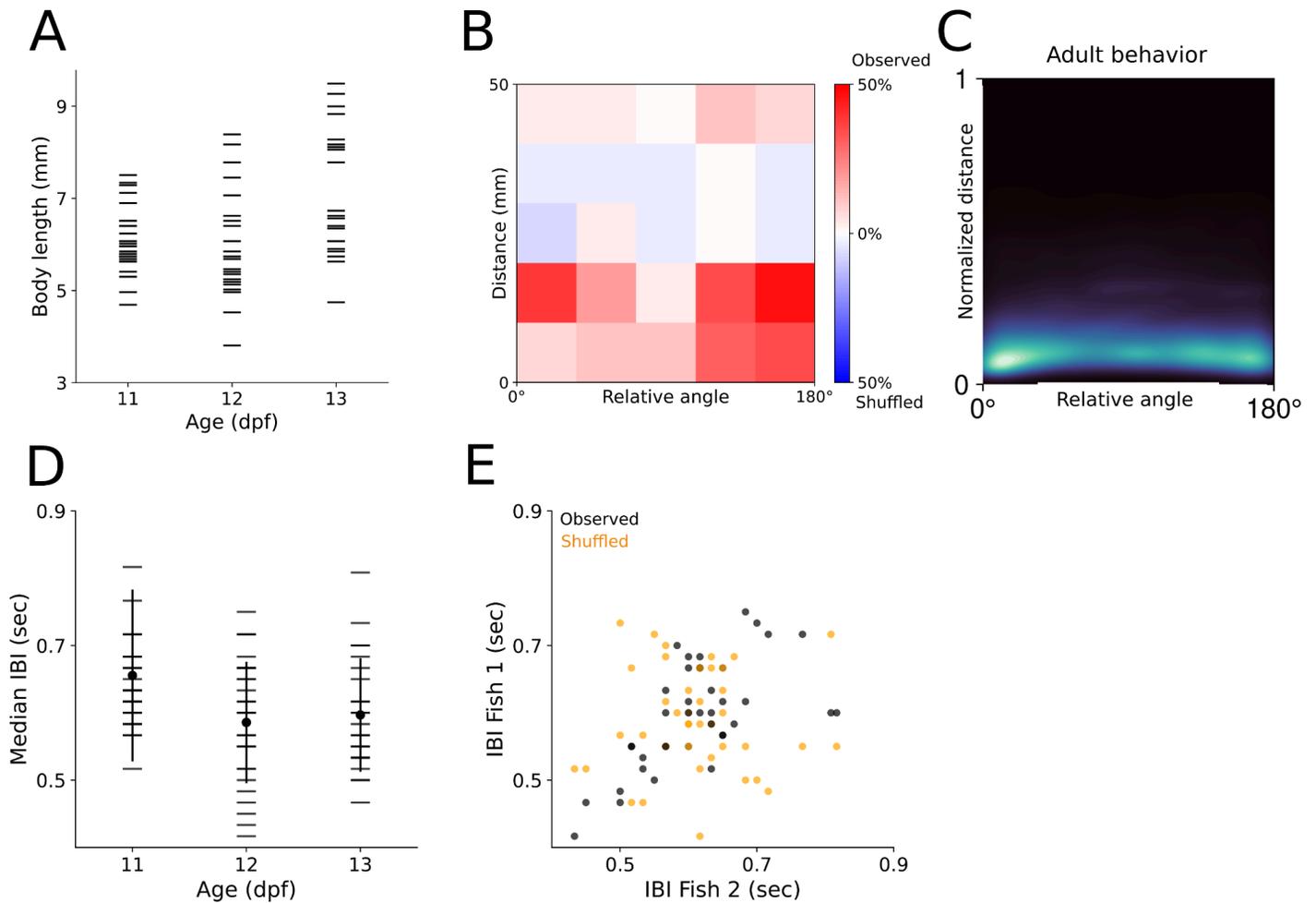

**Figure S1. A.)** Body length by age. **B.)** Statistical quantification of 2-dimensional histograms of distance and relative angle. Colors represent the percentage of pairs that are greater than the 95th percentile of the opposing (observed/shuffled) distribution, showing a significant bias in the observed data towards close and parallel/anti-aligned configurations **C.)** 2D histogram of distance and relative angle for adult pairs of zebrafish. **D.)** Median interbout interval as a function of age. **E.)** Median interbout intervals for each fish plotted against the median interbout interval of their partner. Shuffled data is presented in orange. Linear correlations for observed ($R^2$ = .321, p < 0.001) and shuffled ($R^2$ = .002, p = 0.739).

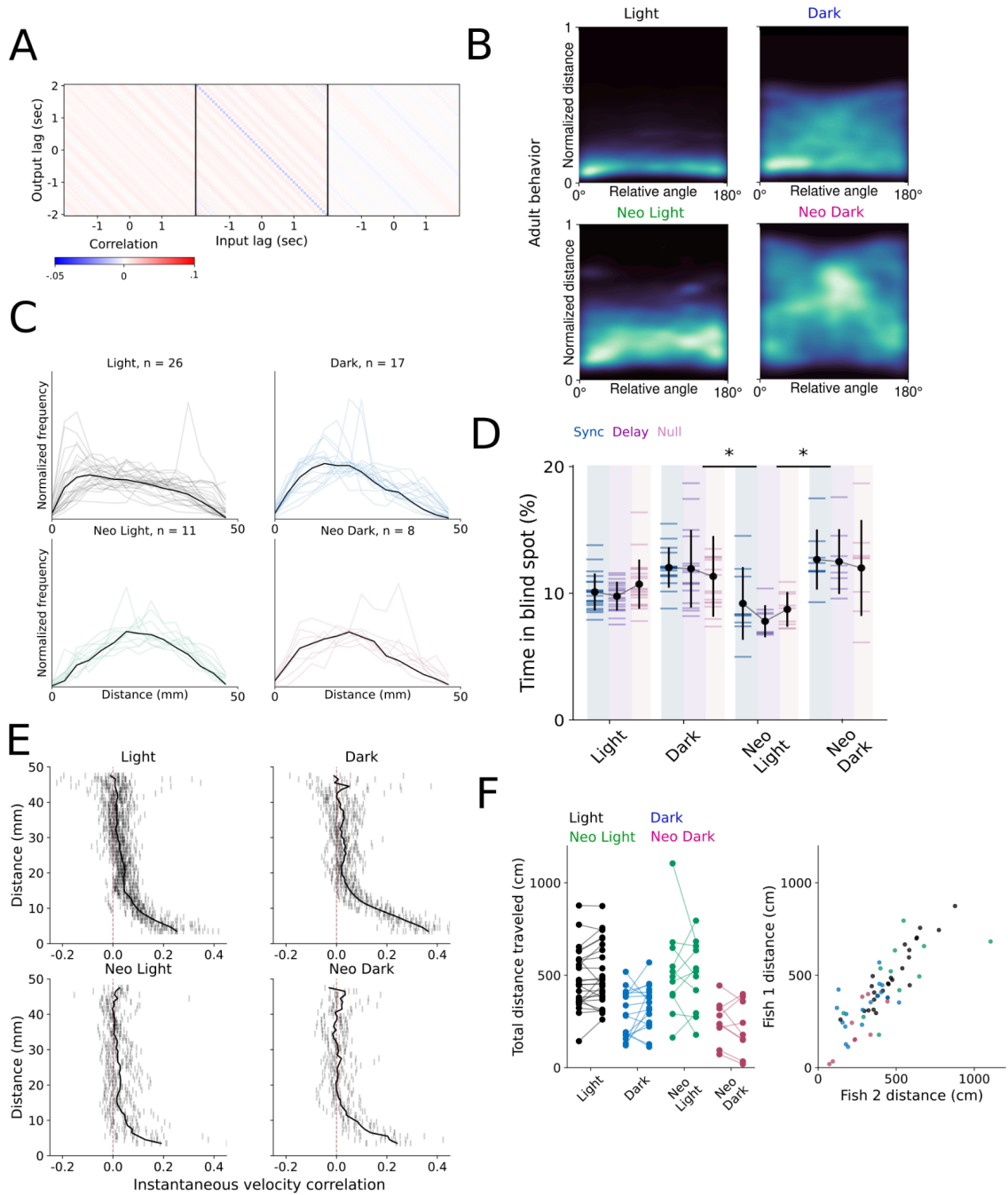

**Figure S2. A.)** Inferred weights from shuffled dataset. **B.)** 2D histograms of adult zebrafish in all sensory conditions. **C.)** Distance histograms of juvenile zebrafish. **D.)** Percentage of time spent in blind spot across conditions. **E.)** Instantaneous velocity correlations as a function of distance across sensory conditions. **F.)** Total distance traveled across sensory conditions, with lines connecting individuals in the same experiment, and total distance traveled for an individual plotted against the distance traveled by its partner. Asterisks indicate a p value below an alpha of 0.05 corrected for multiple comparisons with a Šidák adjustment.

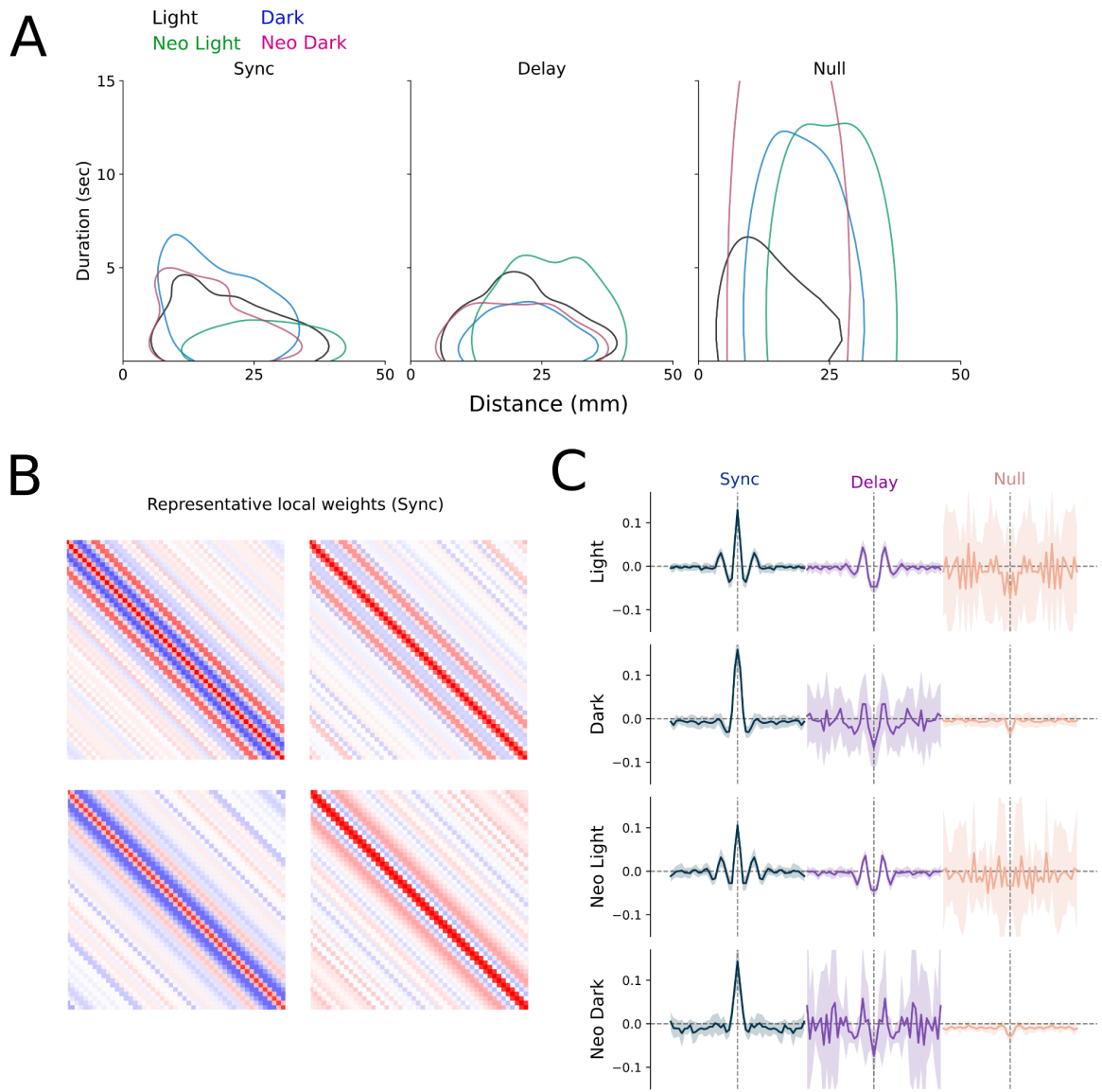

**Figure S3. A.)** Distributions of state duration by distance. **B.)** Representative local inferred weights for the synchronized state in 4 light pairs. **C.)** Average central row of the local inferred weights matrices for all sensory conditions.